\documentclass{article}

\usepackage{arxiv}

\usepackage[utf8]{inputenc} 
\usepackage[T1]{fontenc}    
\usepackage{hyperref}       
\usepackage{url}            
\usepackage{booktabs}       
\usepackage{amsfonts}       
\usepackage{nicefrac}       
\usepackage{microtype}      
\usepackage{lipsum}
\usepackage{graphicx}
\graphicspath{ {./images/} }

\usepackage{caption}
\usepackage{subcaption}
\usepackage{xcolor}

\usepackage[linesnumbered,ruled,vlined]{algorithm2e}

\SetCommentSty{mycommfont}

\usepackage{amsmath,amssymb,amsfonts,amsthm}

\title{Online Heatmap Generation with Both High and Low Weights}

\author{
 Yan Y. Liu \thanks{This manuscript has been authored in part by UT-Battelle, LLC, under contract DE-AC05-00OR22725 with the US Department of Energy (DOE). The US government retains and the publisher, by accepting the article for publication, acknowledges that the US government retains a nonexclusive, paid-up, irrevocable, worldwide license to publish or reproduce the published form of this manuscript, or allow others to do so, for US government purposes. DOE will provide public access to these results of federally sponsored research in accordance with the DOE Public Access Plan (http://energy.gov/downloads/doe-public-access-plan).} \\
  Computational Sciences and Engineering Division\\
  Oak Ridge National Laboratory\\
  Oak Ridge, TN 37831 \\
  \texttt{yanliu@ornl.gov} \\
   \And
 Melissa Allen-Dumas \\
  Computational Sciences and Engineering Division\\
  Oak Ridge National Laboratory\\
  Oak Ridge, TN 37831 \\
  \texttt{allenmr@ornl.gov} \\
}

\begin{document}
\maketitle
\begin{abstract}
Heatmap is a common geovisualization method that interpolates and visualizes a set of point observations on a map surface. Most of online web mapping libraries implement a one-pass heatmap algorithm using HTML5 canvas or WebGL for efficient heatmap generation. However, such implementation applies additive operations that accumulate the rendering of point weights on the map surface grid, making it inappropriate for visualizations that require the highlighting of both low and high weights. We introduce \textit{hilomap}, an online heatmap algorithm that highlights surface areas where points with both low and high trends are located. An HTML5 Canvas-based reference implementation on OpenLayers is presented and evaluated.
\end{abstract}


\section{Introduction}
Spatial heat map, referred to as \textit{heatmap} hereafter, is a common visualization method in geographic information systems. Given a set of point observations, a map surface is rendered to visualize the relative density of these points on each surface grid cell. The color representation of the density measure and its distribution on the map surface help readers understand how the density pattern embedded in the point observations is clustered or varied spatially. There are a wide range of methods available~\cite{esri1} for determining the density of surface grid cells from nearby sample points, from simple point overlay methods, spatial interpolation methods (e.g., inverse distance weighting (IDW)), kernel density estimation (KDE), to sophisticated geostatistical methods such as kriging. 

Heatmap is desirable for visualizing a large number of point observations that spatially overlap. Sparsely distributed observations can be visualized using clustering or other aggregation methods, instead. However, as the number of points increases, the computational cost increases, too. If computationally intensive interpolation algorithms are needed to provide statistically sound results (e.g., using kriging), a typical approach to online heatmap generation conducts costly interpolation computation at server side in cloud or using high-performance computing first. The interpolated map rasters are then rendered and published as an online image service (e.g., via the Open Geospatial Consortium (OGC) Web Coverage Service and the Web Map Service). Users then load rendered map tiles into the map panel within a browser. This approach introduces long turnaround time that may be unacceptable for real-time online mapping. Most of popular web mapping libraries, such as OpenLayers~\cite{ol5heatmap} and Leaflet~\cite{leaflet1,leaflet2}, take a tradeoff and adopt simple but fast point overlay methods.

In this paper, a new heatmap algorithm, \emph{hilomap}, is developed to apply an indirect point overlay method to efficiently process a large number of weighted points and highlight both low and high trends on a map grid, a limitation that exists in current online heatmap libraries. The paper is organized as follows. We introduce the heatmap generation problem formulation and focus on the discussion of point overlay methods used in rendering heatmap as a vector layer in web mapping. Two point overlay methods that are implemented in OpenLayers and Leaflet are described to illustrate how to develop heatmap as a vector layer based on HTML5 canvas, a raster-based graphics framework within a browser. Specifically, necessary interactions between geospatial data points, the map surface, and graphics rendering in HTML5 canvas are presented to illustrate how point weights are transformed to surface weights and how such weights influence the final color rendering of surface cells under spatial overlapping. We will then show the limitations of existing heatmap libraries in visualizing data with low and high extremes, on which trends toward both low and high extremes need to be highlighted. A new heatmap algorithm, called \textit{hilomap}, is proposed to address this issue. A reference implementation in OpenLayers is then illustrated and evaluated.

\section{Algorithms}

\subsection{Problem formulation}

\begin{figure}
  \centering
     \begin{subfigure}[b]{0.462\textwidth}
         \centering
         \textcolor{lightgray}{\frame{\includegraphics[width=\textwidth]{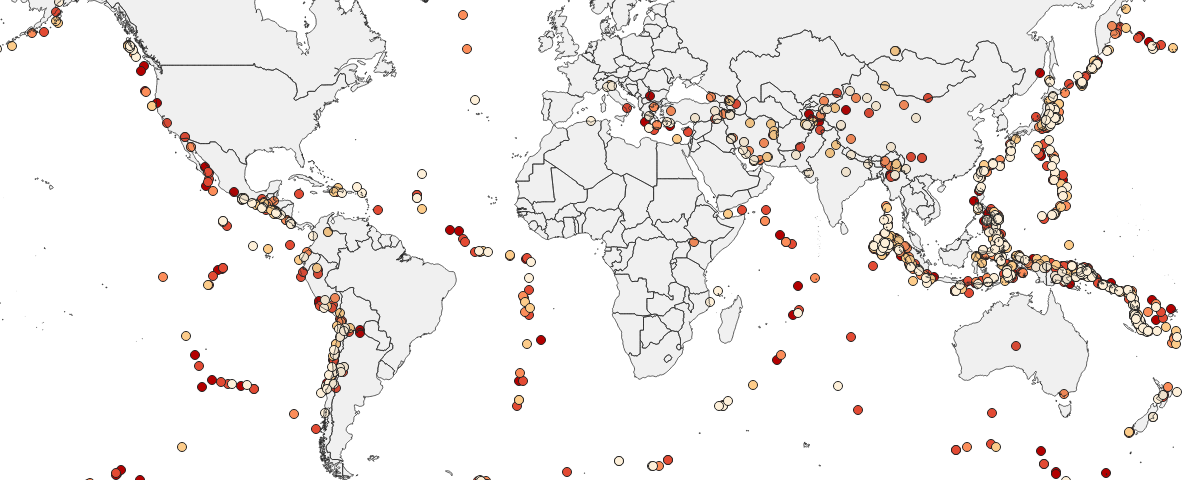}}}
         \caption{Data points}
         \label{fig:olexdata}
     \end{subfigure}
     \hfill
     \begin{subfigure}[b]{0.498\textwidth}
         \centering
         \textcolor{lightgray}{\frame{\includegraphics[width=\textwidth]{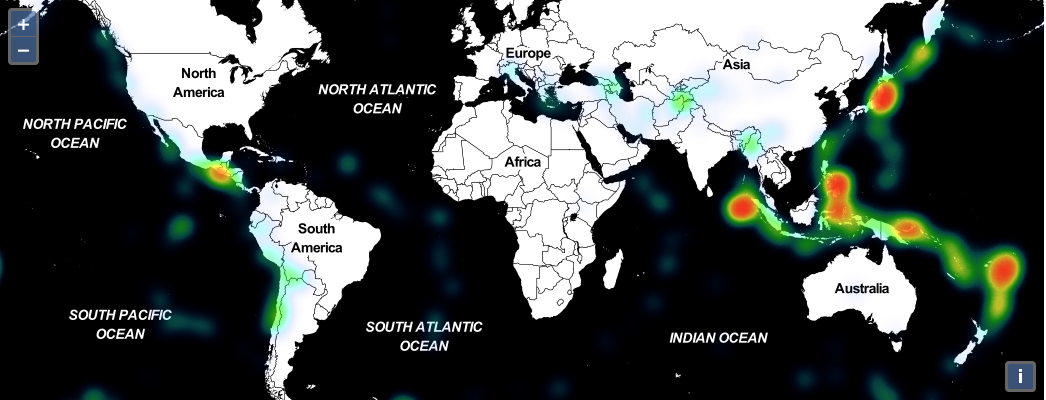}}}
         \caption{Heatmap}
         \label{fig:olex}
     \end{subfigure}
     \par\medskip
     \begin{subfigure}[b]{0.28\textwidth}
         \centering
         \includegraphics[width=\textwidth]{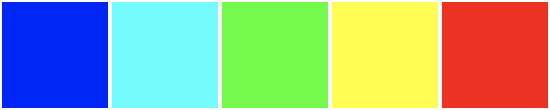}
         \caption{The color gradient, low $\rightarrow$ high}
         \label{fig:olcolorband}
     \end{subfigure}
  \caption{The 2012 global earthquake data and its heatmap, used in OpenLayers examples. }
  \label{fig:olsample}
\end{figure}

Given $k$ point observations $D = \{ (p_i, w_i) \}$, $i = 1 \cdots k$, where $p_i$ is the point location and $w_i$ is the point weight, a heatmap generation algorithm creates a map surface grid $G_M$ with $m$ rows and $n$ columns, on which each grid cell/pixel is rendered with a color and an opacity that reflect the relative density derived from point weights. Figure~\ref{fig:olsample} illustrates an example heatmap layer in OpenLayers v5.3.3 (Figure~\ref{fig:olex}) derived from a global earthquake observation dataset (Figure~\ref{fig:olexdata}). While the point dataset shows the location and the weight (i.e., magnitude) of individual earthquake events, the heatmap presents a spatial pattern on the projected global map surface by considering the spatial distribution of points and the influence of weights. 

\begin{algorithm}[H]
\DontPrintSemicolon
\caption{Pseudo algorithm for general interpolation-based heatmap generation.}
\label{algo:interpo}
\For{$i=0; i<m; i++$} {
  \For{$j=0; j<n; j++$} {
    \tcc{\textit{interpolate()} scans $D$}
    $c_{ij} = interpolate(D)$ \;
    $G_M[i][j] = render(c_{ij})$
  }
}
\end{algorithm} 

To determine the value of each cell on the surface grid $G_M$, an interpolation method may need to consider all the data points, illustrated in Algorithm~\ref{algo:interpo}. The function $interpolate()$ computes a grid cell by scanning the entire point data. For example, IDW computes a cell as $c_{ij} = \frac{\sum_{(p,w) \in D} w / distance(c_{ij}, p)}{\sum_{(p,w) \in D} 1 / distance(c_{ij}, p)}$, such that it is inversely proportional to the distance between the cell and the points. The computational complexity of Algorithm~\ref{algo:interpo} is $O(mnk)$, which is high. Sophisticated spatial statistical methods such as kriging is even more computationally costly as it builds a semivariogram on input points first before interpolating grid cells. 

\subsection{Point overlay methods}
Web mapping libraries often employ point overlay methods, instead of IDW or kriging, to provide fast heatmap generation and acceptable user interaction performance requirements. As its name suggests, a point overlay method, $f: D \rightarrow S$, scans the points in $D$ one after another only once and creates a set of rendering shapes (often as circles) $S$. Since this is a one-pass algorithm, it greatly accelerates heatmap generation. There are two types of point overlay methods: direct or indirect. A direct point overlay methods directly draws the points in $D$ onto $G_M$, therefore, $|D| = |S|$. Indirect point overlay methods often assume a heatmap grid $G_H$ that is coarser than the map surface grid $G_M$. Each point falls into a cell on $G_H$. If a cell contains multiple points, an aggregation strategy is applied to select a representative point for the cell. The number of cells on $G_H$ is often much smaller than the number of points in order to accelerate heatmap rendering. 

\begin{figure}
  \centering
     \hspace{1.5cm}
     \begin{subfigure}[b]{0.1\textwidth}
         \centering
         \textcolor{lightgray}{\frame{\includegraphics[width=\textwidth]{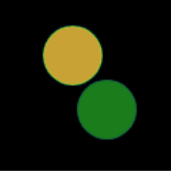}}}
         \caption{radius=4, no blur}
         \label{fig:po1}
     \end{subfigure}
     \hfill
     \begin{subfigure}[b]{0.1\textwidth}
         \centering
         \textcolor{lightgray}{\frame{\includegraphics[width=\textwidth]{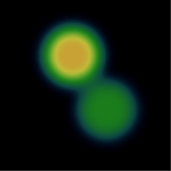}}}
         \caption{radius=4, blurred}
         \label{fig:po2}
     \end{subfigure}
     \hfill
     \begin{subfigure}[b]{0.1\textwidth}
         \centering
         \textcolor{lightgray}{\frame{\includegraphics[width=\textwidth]{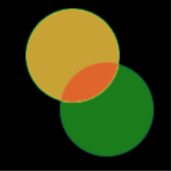}}}
         \caption{radius=9, no blur}
         \label{fig:po3}
     \end{subfigure}
     \hfill
     \begin{subfigure}[b]{0.1\textwidth}
         \centering
         \textcolor{lightgray}{\frame{\includegraphics[width=\textwidth]{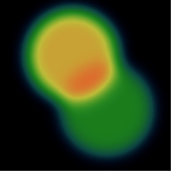}}}
         \caption{radius=9, blurred}
         \label{fig:po4}
     \end{subfigure}
     \hspace{1.5cm}
  \caption{Point overlay rendering.  } 
  \label{fig:po}
\end{figure}

The HTML5 canvas provides two features that make it possible for point overlay methods to work. First, it provides vector shape drawing methods with customized geometry, coloring, and opacity style configuration. Second, if a pixel on the canvas grid is part of the intersection of multiple shapes, the color and the opacity of the pixel are accumulated from all shape drawings. In an HTML5 canvas implementation, a direct point overlay method achieves the effect of spatial overlapping of weights by controlling the cumulative rendering of $S$. A single circle drawing is controlled by a parameter tuple <\emph{radius, blur, opacity}> that can be configured by users. The radius parameter controls the range of spatial diffusion of a point. The blur parameter controls the final color smoothing from the circle center to the boundary of the circle by applying a Gaussian 2D blur operation. The opacity ranges from 0 (transparent) to 1 (opaque). Most of point overlay methods leverage the opacity attribute to carry and accumulate point weights, which is critical for the final rendering of spatial overlapping. When circles do not overlap, the blur operation itself can show the effect of spatial weight decay in a single circle. Figure~\ref{fig:po1} shows two non-overlapping circles with different weights. When blurred, as shown in Figure~\ref{fig:po2}, since the weight within the circle is always non-negative and the boundary has zero weight, the coloring transitions from high in the center to low on the boundary by following the color gradient in Figure~\ref{fig:olcolorband}. 

A direct point overlay rendering process has two steps. First, circles in $S$ are drawn onto $G_M$ without color. $G_M$ is an HTML5 canvas of the map panel, which is a 2D image buffer with each pixel defined by four channels: R (red), G (green), B (blue), and A (alpha, i.e., opacity). Each channel is one byte with value range 0-255. Drawing a circle without color only transforms a point's weight to an opacity value on the A channel of each pixel covered by the circle. Drawings of other overlapping circles will add new opacity values, achieving weight summation. Second, the actual color rendering occurs. The rendered image data of $G_M$ in the first step is fetched for each pixel's alpha value, based on which the actual color is interpolated from a predefined color gradient and the pixel's RGB value is assigned accordingly. In Figure~\ref{fig:po3}, the cumulative effect on opacity in the overlapped area of the two circles renders a higher color than either of the point's. The blur operation helps color blending similarly as in the non-overlapping scenario, as shown in Figure~\ref{fig:po4}.

\begin{figure}
  \centering
  \includegraphics[width=6.4in]{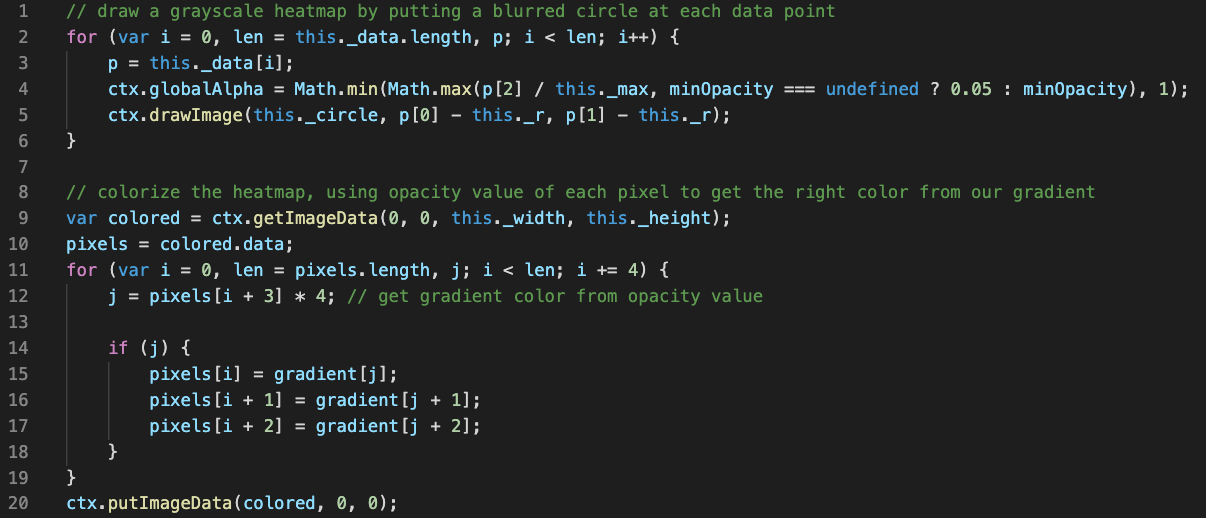}
  \caption{The direct point overlay code snippet in Leaflet's \emph{simpleheat} (edited for formatting). Line 2-6 draws all the point circles, with circle opacity proportional to point weight (a point $p$ is an array of 3 elements: x, y, weight). Line 9-19 fetches pixel-level opacity value accumulated from the previous step, finds the corresponding color on the gradient table, and fills the actual color of each pixel on the map canvas $ctx$. The color gradient is a continuous byte array of size $256 \times 3$. It serves as a lookup table that maps an alpha channel value (i.e., the opacity value) between 0-255 to a 3-byte RGB value.}
  \label{code:simpleheat}
\end{figure}

Direct point overlay is widely supported in web mapping libraries, such as Leaflet's $simpleheat$~\cite{leaflet1} and OpenLayers' Heatmap layer in version 5.3.3~\cite{ol5heatmap}. A JavaScript code snippet of $simpleheat$ is shown in Figure~\ref{code:simpleheat}.

Direct point overlay is simple, but has two limitations. First, the number of rendering operations depends on the number of points $k$, regardless of the resolution and dimensions of the map panel on user's screen. Rendering a large number of points will cause huge energy cost on graphics card and long time delay. Second, the way that weights are accumulated is fixed, giving limited flexibility for other weight aggregation methods. To address these issues, indirect point overlay methods insert a heatmap surface grid $G_H$ of dimension $m' \times n'$ between $D$ and $G_M$, where $m' \leq m$ and $n' \leq n$, shown in Algorithm~\ref{algo:indirect}. The one-pass point scan now fills the content of $G_H$. Only cells on $G_H$ with meaningful content are sent for the drawing and rendering step, which calls a direct point overlay algorithm directly because a cell is represented as an aggregated point.

\begin{algorithm}[H]
\DontPrintSemicolon
\caption{Pseudo algorithm for indirect point overlay.}
\label{algo:indirect}
\tcc{input: $D$ - points; $G_M$ - map surface; $m', n'$ - $G_H$ dimension}
\tcc{Step 1: aggregate points onto $G_H$}
\For{$(p, w) \in D$} {
  $(cx, cy) = coordTransform(p, G_H)$ \tcp*{transform point coordinates to cell coordinates} 
  adjust cell representation $c: (p_c, w_c)$ with $(p, w)$ 
}
$D' = \emptyset$ \;
\For{$c: (p_c, w_c) \in G_H$} {
  \If{$w_c > 0$} {
    $D' = D' \cup \{c\}$ \;
  }
}
\tcc{Step 2: call direct point overlay to generate the heatmap}
$directPointOverlay(D', G_M)$
\end{algorithm} 

\begin{figure}
  \centering
  \includegraphics[width=5in]{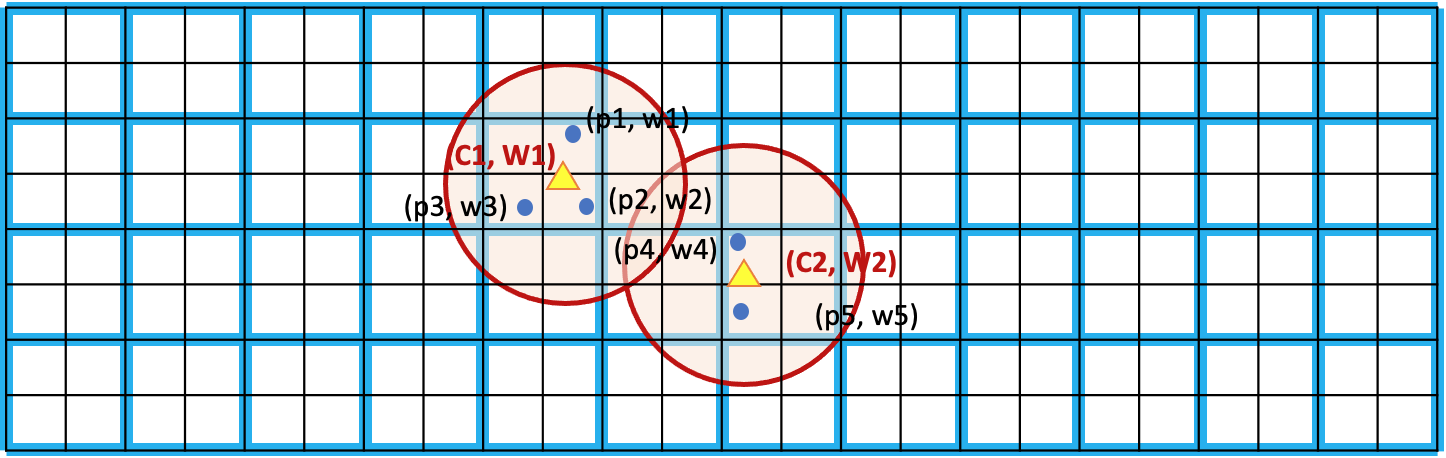}
  \caption{One-pass point scan onto a heatmap surface grid $G_H$ of a coarsened resolution (i.e., the blue grid). The HTML5 canvas grid for the map $G_M$ is a finer grid and is colored black. Points that fall into a blue cell are aggregated into a cell representation, which is also a point with the circle shape. Each non-empty cell selects a representative point with an aggregated weight for actual drawing and rendering.}
  \label{fig:indirect}
\end{figure}

In Algorithm~\ref{algo:indirect}, the computational complexity of the point scan (line 1-3) is $O(k)$. The complexity of  the point rendering (line 4-8) is $O(m'n')$, independent of $k$. When $k \gg m'n'$, the complexity of rendering is still $O(m'n')$. Thus, the performance of rendering can be significantly improved. Furthermore, there are multiple choices for cell presentation. For example, in Figure~\ref{fig:indirect}, a cell is represented by a point whose coordinates, denoted as yellow triangles, are averaged along both x and y directions and the weight can be flexibly determined, e.g., using iterative implementation of aggregation functions such as $sum(), min(), max(), avg()$. Other cell presentations include a representative point for a cell or the cell centroid. Therefore, $G_H$ is a logic surface grid that bridges $D$ and $G_M$. 

\begin{figure}
  \centering
  \includegraphics[width=4.5in]{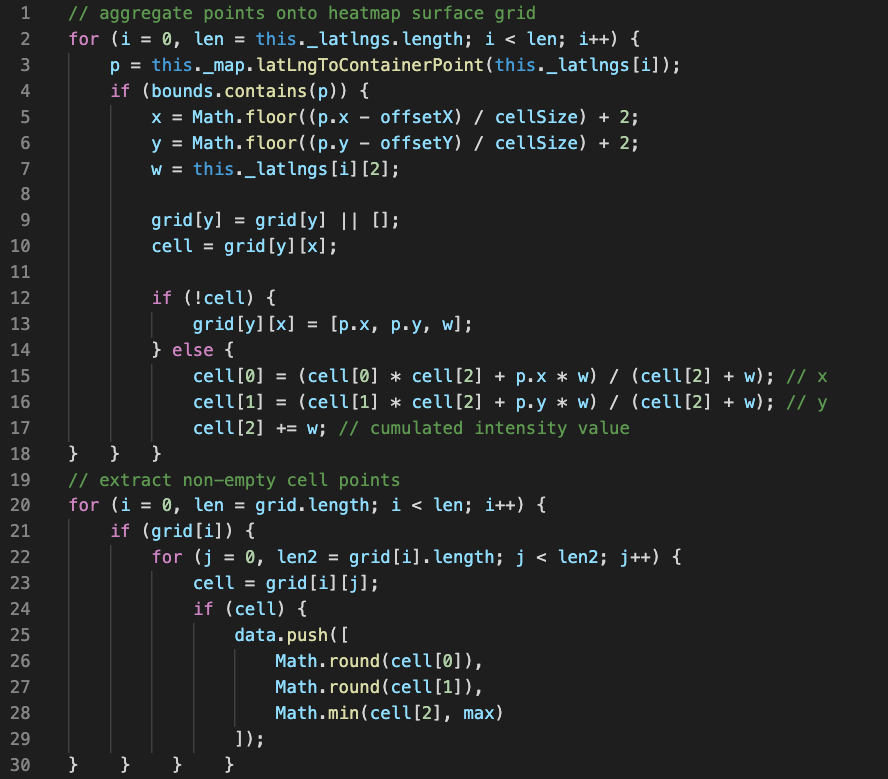}
  \caption{$Leaflet.heat$ implements an indirect point overlay method. Line 2-18 aggregates points onto a virtual heatmap grid. Each cell on this grid is represented as a point. The coordinates of the point are calculated as a weighted average of the coordinates of points that fall into this cell, i.e., $x\_coord(c) = \frac{\sum_{p \text{ within } c} {p_x \times w_p}}{\sum_{p \text{ within } c} {p_x }}$. The weight of the point takes the sum of those data points. Line 20-30 pushes representative points of non-empty cells into a point array for direct point overlay rendering.}
  \label{code:indirect}
\end{figure}

$Leaflet.heat$ in Leaflet is such an implementation using the cell representation in Figure~\ref{fig:indirect}. Figure~\ref{code:indirect} is the relevant code snippet.

\section{Hilomap}

\begin{figure}
  \centering
  \includegraphics[width=3.8in]{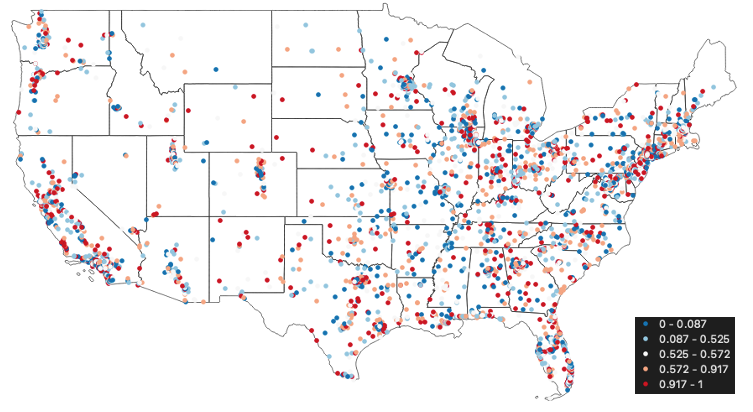}
  \caption{A synthetic data sample on 3387 major U.S. cities. 1/3 data points have low weights, following the Normal distribution with mean=0.1, var=0.03. 1/3 data points have high weights, following the Normal distribution with mean=0.9, var=0.03. The rest of the 1/3 data points follow the Normal distribution with mean=0.5, var=0.03. Weight values are scaled in range [0..1].}
  \label{fig:uscities}
\end{figure}

Aforementioned point overlay algorithms are efficient for visualizing clustered points with a low-to-high trend on weights. An area on the map with clustered points of high weights is highlighted with high color on a color gradient and a sparse area with low weights is rendered with low color on the color gradient. However, when two circles overlap, one with a high weight and the other with a low weight, the overlapped area is rendered using a color gradient that is the sum of the two, which is greater than the weight of either circle, as shown in Figure~\ref{fig:po3}. This creates a limitation that prevents its use in some common visualization scenarios. For example, when an analyst wants to study population change in major cities in the United States, population decreases may be as important as population increases, illustrated in a synthetic data in Figure~\ref{fig:uscities}. When we look at climate change measures, such as temperature, extremely low and extremely high observations need more attention than average ones. When a conventional heatmap is generated from such data, however, low extremes are overridden by high extremes due to the cumulative density effect. For visualizing such data with both low and high extremes, there are two requirements that a conventional heatmap does not satisfy: 1) On a color gradient that changes from low to neutral to high, areas with both low weights and high weights should be highlighted; and 2) When a circle with a high weight overlaps with another one with a low weight, the aggregated weight should neutralize and the coloring should stay around the middle of the color gradient, instead of toward the high end.

\begin{figure}
  \centering
     \begin{subfigure}[b]{0.48\textwidth}
         \centering
         \textcolor{lightgray}{\frame{\includegraphics[width=\textwidth]{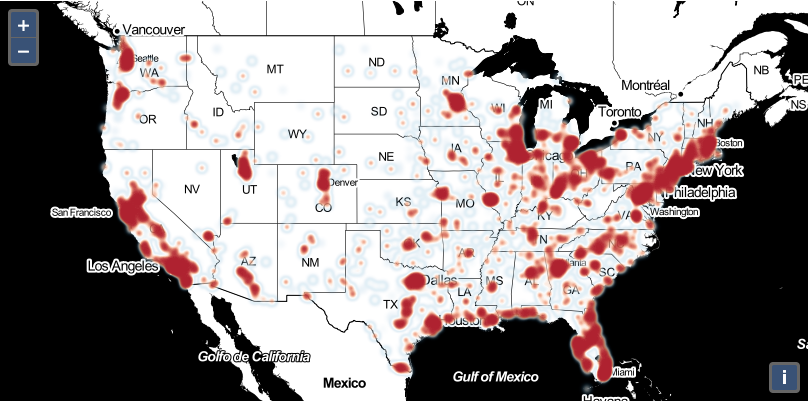}}}
         \caption{OpenLayers v5.3.3 heatmap, radius=4, blur=4}
         \label{fig:heat_r4b4}
     \end{subfigure}
     \hfill
     \begin{subfigure}[b]{0.48\textwidth}
         \centering
         \textcolor{lightgray}{\frame{\includegraphics[width=\textwidth]{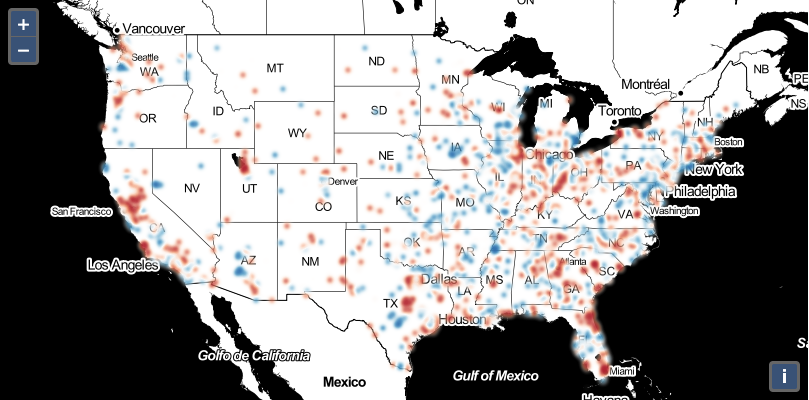}}}
         \caption{hilomap, radius=4, blur=4}
         \label{fig:hilo_r4b4}
     \end{subfigure}
     \par\bigskip
     \begin{subfigure}[b]{0.48\textwidth}
         \centering
         \textcolor{lightgray}{\frame{\includegraphics[width=\textwidth]{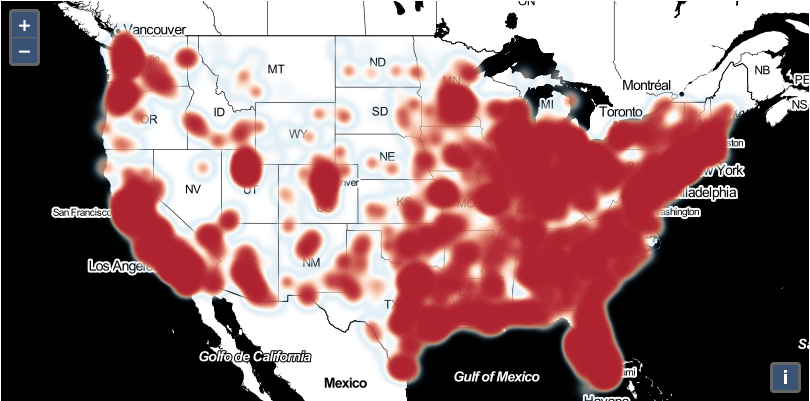}}}
         \caption{OpenLayers v5.3.3 heatmap, radius=10, blur=10}
         \label{fig:heat_r10b10}
     \end{subfigure}
     \hfill
     \begin{subfigure}[b]{0.48\textwidth}
         \centering
         \textcolor{lightgray}{\frame{\includegraphics[width=\textwidth]{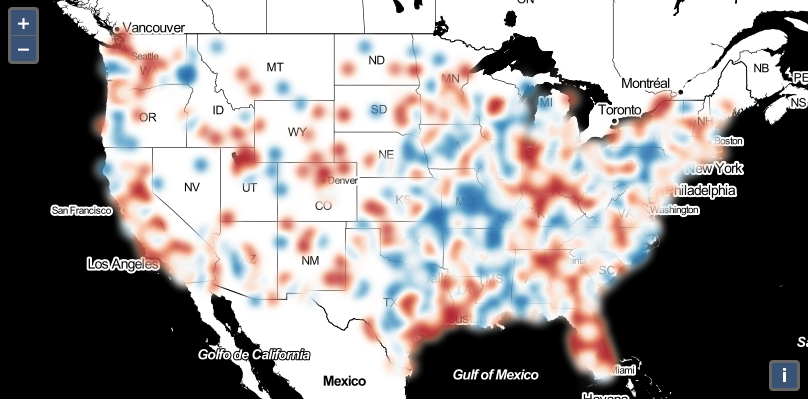}}}
         \caption{hilomap, radius=10, blur=10}
         \label{fig:hilo_r10b10}
     \end{subfigure}
     \par\bigskip
     \begin{subfigure}[b]{0.4\textwidth}
         \centering
         \includegraphics[width=\textwidth]{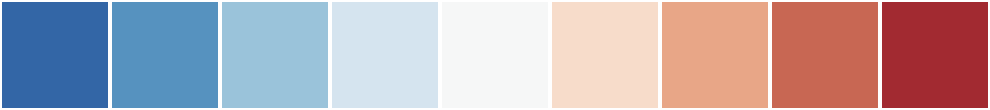}
         \caption{The color gradient, low $\rightarrow$ neutral $\rightarrow$ high}
         \label{fig:color}
     \end{subfigure}
  \caption{Comparison of heatmap and hilomap. }
  \label{fig:maps}
\end{figure}

Figure~\ref{fig:heat_r4b4} and Figure~\ref{fig:heat_r10b10} illustrates the ill-suited heatmap effect. Both heatmaps are dominated by high extremes. The cumulative weight effect is worse with larger radius as spatial overlapping is more likely. The blur operation at the boundary of overlapping areas does not help to show low extremes because low weights are carried in the opacity channel of $G_M$, making a low-extreme area more transparent. One may suggest to create two heatmap layers, one for high and the other for low (using an inverted color gradient from neutral to low), and load them in a map. However, this option is ineffective because of the visual interference to the bottom layer from the opacity of the top layer.

\begin{algorithm}[H]
\DontPrintSemicolon
\caption{Pseudo algorithm for $hilomap$.}
\label{algo:hilomap}
\tcc{input: $D$ - points; $G_M$ - map surface; $m', n'$ - $G_H$ dimension}
\tcc{Step 1: aggregate points onto $G_H$}
\For{$(p, w) \in D$} {
  $(cx, cy) = coordTransform(p, G_H)$ \tcp*{convert point coordinates to cell coordinates} 
  \If{$|w - 0.5| > |w_c - 0.5|$} { 
    $c_x = p_x; c_y = p_y; w_c = w;$ \tcp*{select the point w/ weight furthest to 0.5, the neutral weight}
  } 
}
$D' = \emptyset$ \;
\For{$c: (p_c, w_c) \in G_H$} {
  \If{$c \neq \emptyset$} {
    $D' = D' \cup \{c\}$ \;
  }
}
\tcc{Step 2: render cell points and get low and high pixel weights}
draw circles for cell points with $w_c \leq 0.5$ \;
$lowImg = G_M$ \tcp*{record pixel opacity}
draw circles for cell points with $w_c > 0.5$ \;
$highImg = G_M$ \tcp*{record pixel opacity}
draw all cell points \;
\tcc{Step 3: coloring}
\For{$pixel \in G_M$} {
  reconstruct opacity from $lowImg$ and $highImg$ \;
  fetch color gradient and set RGB values
}
\end{algorithm}

\begin{figure}
  \centering
  \includegraphics[width=5in]{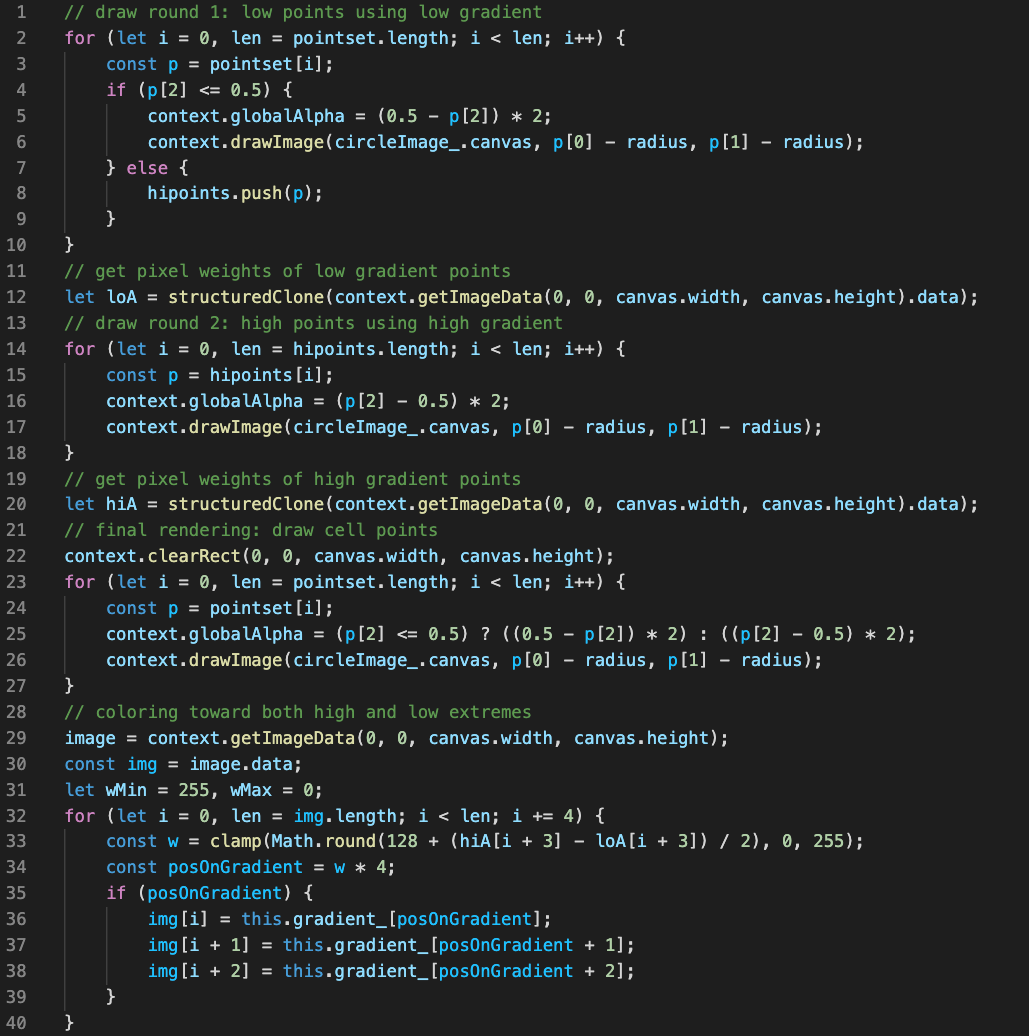}
  \caption{The color rendering code of hilomap.}
  \label{code:hilomap}
\end{figure}

This issue is addressed in $hilomap$, a point overlay method that shows both low and high trends as a single vector layer. The major challenge is in color rendering. In rendering, point weight is scaled between 0 and 1 to fit the value range of the opacity in shape drawing. After the shape is drawn, the opacity value is translated to the alpha channel value, which is one byte and ranges from 0 to 255. The weight-opacity-alpha translation is necessary because: 1) point overlay methods leverage the fact that drawing overlapping shapes provides the sum operation on opacity values; and 2) correct image pixel coloring from a user-defined color gradient can only occur at the final stage by using the alpha channel value to look up the correct color mapping. In conventional heatmap rendering, the scaling to opacity and alpha channel is linearly proportional to the weight value. However, in hilomap, highlighting low extremes means two low opacity values on a pixel should lead to a lower opacity value, which is not possible using the default drawing method. 

Hilomap is an indirect point overlay algorithm that addresses the rendering challenge and satisfy the requirements on visualizing change data, as shown in Algorithm~\ref{algo:hilomap} and the detailed color rendering code snippet (Figure~\ref{code:hilomap}). For cell representation (line 3-4 in Algorithm~\ref{algo:hilomap}), a representative point is chosen as the point whose weight is furthest from the neutral value 0.5. This strategy prefers points on the two ends in the weight range. The weight value is directly copied. Other strategies such as weight averaging also work. If a user requires the color gradient to be consistent with the underlying weight distribution, the neutral value may be set as the statistical mean and the color gradient may be composed to reflect the distribution, e.g. using quantile-based color buckets.

After representative points are determined on $G_H$, circles are drawn to highlight both low and high trends in two steps (line 9-12 in Algorithm~\ref{algo:hilomap}). The first step draws the lower set of points whose weight value is less than 0.5. This is done by turning the weight into negative and scaling back to $[0..1]$ as the opacity value, i.e., $opacity_L = (0.5 - w) \times 2$. This way, low-low point overlay increases the alpha channel value with the upper bound 255. The second step takes the upper set of points on $G_H$, sets the opacity value, i.e., $opacity_H = (w - 0.5) * 2$, and draws them onto canvas. The two steps create two image buffers, $L$ and $H$, that accumulate low and high trends in each buffer's alpha channel, respectively. The difference is that in $L$, a higher alpha value means lower trend and in $H$, it is the opposite. The point drawing details are described in the code snippet in Figure~\ref{code:hilomap} (line 2-20). An extra point drawing is applied to all points in order to set the correct global alpha channel value for each pixel (line 22-27). 

The final coloring step takes the alpha channel data in $L$ and $H$ as input to interpolate the correct color for each pixel from the color gradient. The alpha value indicates the distance to the neutral color and ranges from 0 to 255. Using the formula $w_{color} = 128 + \frac{\alpha_H - \alpha_L}{2}$, a color at the lower end of the color gradient is assigned to pixels with dominant low trend and a color at the higher end is assigned to pixels dominant high trend. In addition, a neutral color is assigned when the alpha values from $L$ and $H$ are similar. The implementation details are sketched in the code snippet in Figure~\ref{code:hilomap} (line 29-40). 

In summary, the hilomap algorithm uses three drawings and one color rendering. The computational complexity of point drawing and coloring is still $O(m'n')$ on $G_H$ of size $m' \times n'$. The three drawings and the memory requirement for keeping two additional image buffers for $L$ and $H$ are more costly than a conventional indirect point overlay method, but in practice, such visualization is fast on modern graphics cards and the major delay is from transferring data points to browser over internet.

Figure~\ref{fig:hilo_r4b4} and Figure~\ref{fig:hilo_r10b10} demonstrate the visualization of population increases and decreases in the synthetic data using hilomap. They clearly show both low and high trends, as well as neutral areas, compared with the OpenLayers v5.3.3 heatmap implementation.

\section{Discussion}

We present algorithmic and technical details of three different point overlay methods for online heatmap generation. The hilomap algorithm aims to address the limitations in the previous two methods in visualizing data where highlighting both low and high trends is important, such as change data where both negative and positive changes need attention. The first two algorithms are demonstrated using Leaflet and OpenLayers, two popular open source web mapping libraries. The reference implementation of the hilomap~\cite{hilomap} is based on OpenLayers v5.3.3. 

HTML5 canvas is assumed to be the map rendering framework. However, WebGL-based heatmaps can also be implemented. In WebGL, {\em vertex shader} is provided to define the positions of points and the {\em fragment shader} is to draw a shape and color. Hilomap implementation in Leaflet and OpenLayers using WebGL will be our future work.

In our illustrations, color gradients transit from low to high in a linear fashion with equal interval. Given a particular weight distribution, a color gradient may be created using other methods such as quantile-based.

\section*{Acknowledgements}

This work is in part supported by the Laboratory Directed Research and Development Program of Oak Ridge National Laboratory (ORNL), managed by UT-Battelle, LLC, for the US Department of Energy under contract DE-AC05-00OR22725. 

\section*{Appendix: Hilomap Code Access}

Hilomap has as an open source JavaScript implementation and can be downloaded at https://github.com/hohe12ly/hilomap. The following are \texttt{OpenLayers} and \texttt{hilomap} installation steps for testing the library.

Install \texttt{nodejs} in conda:

\begin{verbatim}
conda create -yn nodejs python=2.7.18 nodejs
conda activate nodejs
\end{verbatim}

Download OpenLayers v5.3.3 from \texttt{https://github.com/openlayers/openlayers/releases/tag/v5.3.3}.

On Mac OS, do the following to make \texttt{npm} work:

\begin{verbatim}
export PATH=/usr/bin:$PATH
export LD_LIBRARY_PATH=/usr/lib:$LD_LIBRARY_PATH
\end{verbatim}

Go to OpenLayers directory, and:

\begin{verbatim}
npm install
\end{verbatim}

Now install \texttt{hilomap}. Suppose HILODIR is where you download hilomap:

\begin{verbatim}
# copy source files and examples
cp HILODIR/src/ol/layer.js src/ol/ # add hilomap to layer list
cp HILODIR/src/ol/layer/Hilomap.js src/ol/layer/
cp HILODIR/examples/hilomap.* examples/
cp HILODIR/examples/data/kml/uscities_sample.kml examples/
# compile
npm run build-package && npm run build-examples
\end{verbatim}

The hilomap example can then be loaded in browser: \texttt{https://HOST/OLPATH/build/examples/hilomap.html}, where HOST is the web server host name (e.g., \texttt{localhost}) and OLPATH is the web path to OpenLayers deployment on the web server.


\bibliographystyle{unsrt}  


\end{document}